\documentclass[aps,prb,twocolumn,groupedaddress,showpacs, floatfix]{revtex4}
\usepackage{graphicx,amsmath}

\begin{document}
\title{Lattice susceptibility for 2D Hubbard Model within dual fermion method}

\author{Gang Li}
\author{Hunpyo Lee}
\author{Hartmut Monien}
\affiliation{Physikalisches Insititut, Universit\"at Bonn, 53115 Bonn, Germany}
\date{\today}

\begin{abstract}
  In this paper, we present details of the dual fermion (DF) method to study
  the non-local correction to single site DMFT. The DMFT two-particle Green's
  function is calculated using continuous time quantum monte carlo (CT-QMC)
  method. The momentum dependence of the vertex function is analyzed and its
  renormalization based on the Bethe-Salpeter equation is performed in
  particle-hole channel. We found a magnetic instability in both the dual and
  the lattice fermions. The lattice fermion susceptibility is calculated at
  finite temperature in this method and also in another recently proposed
  method, namely dynamical vertex approximation (D$\Gamma$A). The comparison
  between these two methods are presented in both weak and strong coupling
  region. Compared to the susceptibility from quantum monte carlo (QMC)
  simulation, both of them gave satisfied results.
\end{abstract}
 
\maketitle

\section{Introduction\label{Introduction}}

Strongly correlated electron systems, such as heavy fermion compounds,
high-temperature superconductors, have gained much attention from both
theoretical and experimental point of view. The competition between the
kinetic energy and strong Coulomb interaction of fermions generates a lot
of fascinating phenomena. Various theoretical approaches have been developed
to treat the regime of intermediate coupling. The widely used perturbative
methods, such as random phase approximation (RPA), fluctuation exchange
(FLEX)\cite{Bickers-1989,Bickers-1991}, and the two-particle self-consistent
(TPSC)\cite{Tremblay-1995,Tremblay-1994} method are based on the expansion in
the Coulomb interaction which is only valid in weak-coupling. To go beyond the
perturbative approximation and to gain insight of the correlation effects of
the fermion systems, new theoretical methods are needed. Dynamical mean field
theory (DMFT)\cite{Metzner-1989,Hartmann-1984,George-1996} is a big step
forward in the understanding Metal-Insulator transition. 

Dynamical mean field theory maps a many-body interacting system on a lattice
onto a single impurity embedded in a non-interacting bath. Such a mapping
becomes exact in the limit of infinite coordination number. All local temporal
fluctuations are taken into account in this theory, however spatial
fluctuations are treated on the mean field level. DMFT has been proven a
successful theory describing the basic physics of the Mott-Hubbard
transition. But the non-local correlation effect can't always be
omitted. Although, straight forward extensions of
DMFT\cite{Hettler-1998,Kotliar-2001,Okamoto-2003,Potthoff-2003,Maier-2005} 
have captured the influence of short-range correlation, these methods are
still not capable of describing the collective behavior, e.g. spin wave
excitations of
many-body system. At the same time, most of the numerically exact impurity
solvers require a substantial amount of time to achieve a desired accuracy even
on a small cluster, which makes the investigation of larger lattice to be
impossible.  

Recently, some efforts have been made to take the spatial fluctuations into
account in different ways\cite{Toschi-2007, Rubtsov-2006, Kusunose-2006,
  Tokar-2007, Slezak-2006}. All these methods construct the non-local
contribution of DMFT from the local two-particle vertex. The electron
self-energy is expressed as a function of the two-particle vertex and 
the single-particle propagator. The cluster extention of DMFT considers
the correlation within the small cluster. Compared to these, the 
diagrammatic re-summation technique involved in these new methods makes them
only approximately include the non-local corrections. While, long range
correlations are also considered in these methods and the computational burden
is not serious. 

In this paper we will apply the method of Rubtsov\cite{Rubtsov-2006} to
consider the vertex renormalization of the DF through the Bethe-Salpeter
equation. Lattice susceptibility is calculated from the renormalized DF
vertex.  

The paper is organized as follows: In Sec. \ref{Details} we summarize the
basic idea of the DF method and give details of the calculation. The
DMFT two-particle Green's function and the corresponding vertex calculation
are implemented in CT-QMC in Sec. \ref{Vertex}. The frequency dependent vertex
is modified through the Bethe-Salpeter Equation to obtain the momentum
dependence in Sec. \ref{Mom-Vex}. In Sec. \ref{application} we present the
calculation of the lattice susceptibility and compare it with QMC results and
also the works from Toschi\cite{Toschi-2007}. The conclusions are summarized
in Sec. \ref{conclusion}, where we also present possible application.

\section{The DF method\label{Details}}

We study the general one-band Hubbard model at two dimensions
\begin{equation}
  H=\sum_{k,\sigma}\epsilon_{k,\sigma}c_{k\sigma}^{\dagger}
  c_{k\sigma}^{\phantom\dagger}+U\sum_{i}n_{i\uparrow}n_{i\downarrow} 
\end{equation}
$c_{k\sigma}^{\dagger}(c_{k\sigma})$ creates (annihilates) an electron with
spin-$\sigma$ and momentum $k$. The dispersion relation is
$\epsilon_{k}=-2t\sum_{i=1}^{N}\cos k_{i}$, where $N$ is the number of lattice
sites. The basic idea of the DF method\cite{Rubtsov-2006} is to
transform the hopping between different sites into coupling to an auxiliary 
field $f(f^{\dagger})$. By doing so, each lattice site can be viewed as an
isolated impurity. The interacting lattice problem is reduced to solving a
multi-impurity problem which couples to the auxiliary field. This can be done
using the standard DMFT calculation.  After integrating out the lattice
fermions $c(c^{\dagger})$ one can obtain an effective theory of the auxiliary
field where DMFT serves as an starting point of the expansion over the
coupling between each impurity site with the auxiliary field.  

To explicitly demonstrate the above idea we start from the action of DMFT
which can be written as   
\begin{equation}\label{original_fermion}
  S[c^{+},c]=\sum_{i}S_{imp}^{i}-\sum_{\nu,k,\sigma}(\Delta_{\nu}
  -\epsilon_{k\nu})c_{\nu k\sigma}^{\dagger}c_{\nu k\sigma}^{\phantom\dagger}
\end{equation} 
where $\Delta_{\nu}$ is the hybridization function of the impurity problem
defined by $S_{imp}^{i}$ which is the action of an isolated impurity at site
$i$ with the local Green's function $g_{\nu}$. Using the Gaussian identity, we
decouple the lattice sites into many impurities which couple only to the field
$f$ 
\begin{eqnarray}\label{auxiliary_field}
  S[c^{\dagger},c;f^{\dagger},f]&=&\sum_{i}S_{imp}^{i}+\sum_{k,\nu,\sigma}
  [g_{\nu}^{-1}(c_{k\nu\sigma}^{\dagger}f_{k\nu\sigma}^{\phantom\dagger}+h.c.)
  \nonumber\\
  &&\hspace{1cm}+g_{\nu}^{-2}(\Delta_{\nu}-\epsilon_{k})^{-1}
  f_{k\nu\sigma}^{\dagger}f_{k\nu\sigma}^{\phantom\dagger}] 
\end{eqnarray}
The equivalence of Eqs. (\ref{original_fermion}) and (\ref{auxiliary_field})
form an exact relation between the Green's funtion of the lattice electrons and
the DF.  
\begin{equation}\label{relation}
  G_{\nu,k}=g_{\nu}^{-2}(\Delta_{\nu}-\epsilon_{k})^{-2}
  G_{\nu,k}^{d}+(\Delta_{\nu}-\epsilon_{k})^{-1}
\end{equation}
This relation is easily derived by considering the derivative over
$\epsilon_{k}$ in the two actions. Eq. (\ref{relation}) allows now to solve
the many-body ``lattice'' problem based on DMFT which is different from the
straight forward cluster extension. The problem is now to solve the Green's
function of the DF $G^{d}_{\nu,k}$. It is determined by integrating 
Eq. (\ref{auxiliary_field}) over $c^{\dagger}$ and $c$ yielding a Taylor
expansion series in powers of $f^{\dagger}$ and $f$. The Grassmann integral
ensures that $\bar{f}$ and $f$ appear only in pairs associated with the
lattice fermion n-particle vertex obtained from the single-site DMFT
calculation. In this paper we restrict our considerations to the two-particle
vertex $\gamma^{(4)}$.
    
\begin{figure}[b]
  \includegraphics[width=200pt]{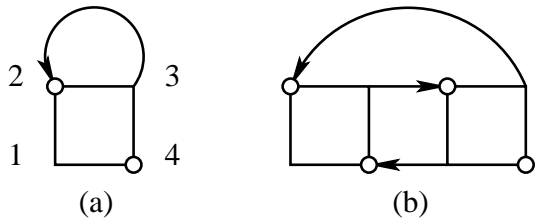}%
  \caption{The first two self-energy diagrams. They are composed of the local
  vertices function and DF propagator. \label{Self-Energy}}
\end{figure}

Expanding the Luttinger-Ward functional in $\gamma^{4}$, the first two
contributions to the self energy function are the diagrams shown in
Fig. \ref{Self-Energy}. Diagram (a) vanishes for the bare DF since this
diagram exactly corresponds to the DMFT self consistency. Therefore the first
non-local contribution is given by diagram (b). The self-energy for these two
diagrams are 
\begin{subequations}
  \begin{align}
    \Sigma^{(1)}_{\sigma}(k_{1}) &= -\frac{T}{N}\sum_{\sigma^{\prime},
      k_{2}}G_{\sigma^{\prime}}^{d}
    (k_{2})\gamma^{(4)}_{\sigma\sigma^{\prime}}
    (\nu,\nu^{\prime};\nu^{\prime},\nu) \label{self-energy1} \\
    \Sigma^{(2)}_{\sigma}(k_{1}) &= -\frac{T^{2}}{2N^{2}}\sum_{2,3,4}
    G^{d}_{\sigma_{2}}(k_{2})G^{d}_{\sigma_{3}}(k_{3})
    G^{d}_{\sigma_{4}}(k_{4})\nonumber\\
    & \gamma^{(4)}_{\sigma_{1234}}(\nu_{1},\nu_{2};\nu_{3},\nu_{4})
    \gamma^{(4)}_{\sigma_{4321}}(\nu_{4},\nu_{3};\nu_{2},\nu_{1})\nonumber\\
    & \delta_{k_{1}+k_{2}, k_{3}+k_{4}}\delta_{\sigma_{1}+\sigma_{2},
      \sigma_{3}+\sigma_{4}} \label{self-energy2}
  \end{align}
\end{subequations}
Here space-time notation is used, $k=(\vec{k},\nu)$,
$q=(\vec{q},\omega)$. Fermionic Matsubara frequency is 
$\nu_{n}=(2n+1)\pi/\beta$, bosonic frequency is $\omega_{m} = 2m\pi/\beta$
where $\beta$ is the inverse temperature. Together with the bare DF
Green's function
$G^{d}_{0}(k)=-g_{\nu}^{2}/[(\Delta_{\nu}-\epsilon_{k})^{-1}+g_{\nu}]$, the
new Green's function can be derived from the Dyson equation 
\begin{equation}\label{Dyson}
  [G^{d}(k)]^{-1}=[G_{0}^{d}(k)]^{-1}-\Sigma^{d}(k)
\end{equation}

The algorithm of the whole calculation is: 
\begin{enumerate}
\item Set initial value of $\Delta_{\nu}$ for the first DMFT loop.
\item Determine the single-site DMFT Green's function $g_{\nu}$ from
  the hybridization function $\Delta_{\nu}$. The self-consistency condition
  ensures that the first diagram of the DF self-energy is very small. 
\item Go through the DMFT loop once again to calculate the two-particle
  Green's function and corresponding $\gamma$-function. The method for
  determining the $\gamma$-function is implemented for both strong and
  weak-coupling CT-QMC in the next section of this paper.  
\item Start an inner loop calculation to determine the DF Green's
  function and in the end the lattice Green's function.
  \begin{enumerate}
  \item From Eqs. (\ref{self-energy1}), (\ref{self-energy2}) and the Dyson
    equation (\ref{Dyson}) to calculate the self-energy of the DF. 
  \item Repeatly use Eq. (\ref{self-energy1}), (\ref{self-energy2}) and
    Eq. (\ref{Dyson}) until the convergence of the DF Green's function
    is achived.  
  \item The lattice Green's function is then given by Eq. (\ref{relation})
    from that of the DF.
  \end{enumerate}
\item Fourier transform the momentum lattice Green's function into real space.
  And from the on-site component $G_{ii}$ to determine a new hybridization
  function $\Delta_{\nu}$ which is given by Eq. (\ref{Old-New}).
\item Go back to the Step 3. and iteratively perform the outer loop until the
  hybridization $\Delta_{\nu}$ doesn't change any more. 
\end{enumerate} 

Although diagram (a) is exactly zero for the bare DF Green's
function, it gives non-zero contribution from the second loop where the DF
Green's function is updated from Eq. (\ref{Dyson}). As a result, the  
hybridization function should also be updated before the next DMFT loop is
performed . This is simply done by setting the local full DF Green's
function to zero, together with the condition that the old hybrization
function forces the bare local DF Green's function to be zero
($\sum_{k}G^{0,d}_{\nu,k}=0$), we obtain a set of equations 
\begin{subequations}
  \begin{align}
    & \frac{1}{N}\sum_{k}[G_{\nu,k} - (\Delta^{New}_{\nu}-\epsilon_{k})^{-1}]
    g_{\nu}^{2}(\Delta^{New}_{\nu}-\epsilon_{k})^{2} = 0 \\
    & \frac{1}{N}\sum_{k}[G^{0}_{\nu,k} -
    (\Delta^{Old}_{\nu}-\epsilon_{k})^{-1}] 
    g_{\nu}^{2}(\Delta^{Old}_{\nu}-\epsilon_{k})^{2} = 0 
  \end{align}
\end{subequations}
which yields 
\begin{equation}
  \Delta_{\nu}^{New}-\Delta_{\nu}^{Old}\approx\frac{1}{N}\sum_{k}(G_{\nu,k}-
  G^{0}_{\nu,k})(\Delta^{Old}_{\nu}-\epsilon_{k})^{2}
\end{equation}
This equation finally gives us the relation between the new and old
hybridization function. 
\begin{equation}\label{Old-New}
  \Delta_{\nu}^{New}=\Delta_{\nu}^{Old}+g_{\nu}^{2}G_{loc}^{d}
\end{equation}

In the whole calculation, the DF perturbation calculation converges
quickly. The most time consuming part of this method is the DMFT calculation
of the two particle Green's function. There are some useful symmetries to
accelerate the calculation. As already pointed out\cite{Abrikosov:QFT,
  Nozieres:1964}, it is convenient to take the symmetric form of the
interaction term. The two particle Green's function is then a fully
antisymmetric function. Such fully antisymmetric form is very useful to speed
up the calculation of the two particle Green's function. One does not need to
calculate all the frequency  points within the cutoff in Mastsubara space, a
few special points are calculated and the values for the other points are
given by that of those special points through antisymmetric property. In the
DF self energy calculation, we always have the convolution type of momentum
summation which is very easy to be calculated by fast fourier transform (FFT). 

\section{CT-QMC and two-particle vertex}\label{Vertex}

From the above analysis, the key idea of the DF method is to
construct the nonlocal contribution from the auxiliary field and the DMFT
two-particle Green's function. Therefore it is quite important to accurately 
determine the two-particle vertex. Here we adapt the newly developed CT-QMC
method\cite{Rubtsov-2005, Werner-2006(1), Werner-2006(2)} to calculate the two
particle Green's function $\chi$.  

First we briefly outline the CT-QMC technique. For more details, we refer the
readers to\cite{Rubtsov-2005, Werner-2006(1), Werner-2006(2)}. Here we discuss
the two-particle Green's function and some numerical implemetations in more
detailed. Two variants of the CT-QMC methods have been proposed based on the
diagrammatic expansion. Unlike the Hirsch-Fye method, these methods don't have
a Trotter error and can approach the low temperature region easily. In the
weak-coupling method\cite{Rubtsov-2005} the non-interacting part of the
partition function is kept and expanded the interaction term into Taylor
series. Wick's theorem ensures that the corresponding expansion can be written
into a determinant at each order  
\begin{equation}
  {\cal Z}=\sum_{k}\frac{(-U)^{k}}{k!}\int d\tau_{1}\cdots d\tau_{k}e^{-S_{0}}
  \det[D_{\uparrow}D_{\downarrow}]
\end{equation} 
with
\begin{equation}
  D_{\uparrow}D_{\downarrow}=
  \left(
    \begin{array}{cc}
      \cdots & G_{\uparrow}(\tau_{1}-\tau_{k}) \cr
      \cdots & \cdots 
    \end{array}
  \right)
  \left(
    \begin{array}{cc}
      \cdots & \cdots \cr
      G_{\downarrow}(\tau_{k}-\tau_{1}) & \cdots
    \end{array}
  \right)
\end{equation}
where $S_{0}$ is the non-interacting action and $G^{0}$ is the Weiss field, 
and the one-particle Green's function is measused as 
\begin{equation}
  G(\nu)=G^{0}(\nu)-\frac{1}{\beta}G^{0}(\nu)\sum_{i,j}M_{i,j}
  e^{i\nu(\tau_{i}-\tau_{j})}G^{0}(\nu)
\end{equation}

In the strong coupling method the effective action is expanded in the
hybridization function by integrating over the non-interacting bath degrees of
freedom. Such an expansion also yields a determinant.  
\begin{eqnarray}
  &&{\cal Z}=TrT_{\tau}e^{-S_{loc}}\prod_{\sigma}
  \sum_{k_{\sigma}}\frac{1}{k_{\sigma}!}\int d\tau_{1}^{s}\cdots
  d\tau_{k_{\sigma}}^{s}\int d\tau_{1}^{e}\cdots d\tau_{k_{\sigma}}^{e}
  \nonumber\\
  &&\Psi_{\sigma}(\tau^{e})\left(
  \begin{array}{ccc}
    \Delta(\tau_{1}^{e}-\tau_{1}^{s}) & \cdots & \Delta(\tau_{1}^{e}
    -\tau_{k_{\sigma}}^{s})\\
    \cdots & \ddots & \cdots \\
    \Delta(\tau_{k_{\sigma}}^{e}-\tau_{1}^{s}) & \cdots & 
    \Delta(\tau_{k_{\sigma}}^{e}-\tau_{k_{\sigma}}^{s})
  \end{array}\right)
  \Psi_{\sigma}^{\dagger}(\tau^{s})
\end{eqnarray}
Here $\Psi(\tau) = (c_{1}(\tau), c_{2}(\tau),\cdots,
c_{k_{\sigma}(\tau)})$. The action is evaluated by a Monte Carlo random walk
in the space of expansion order $k$. Therefore the corresponding hybridization
matrix changes in every Monte Carlo step. One particle Green's function is
measured from the expansion of hybridization function as
$G(\tau_{j}^{e}-\tau_{i}^{s})=M_{i,j}$. $M$ is the inverse matrix of the
hybridization function. Apparently one needs to calculate this inverse matrix
in every update step which is time consuming, fortunately it can be obtained by
the fast-update algorithm\cite{Rubtsov-2005}.   

At the same time such a relation allows direct measurement of the Matsubara
Green's function       
\begin{equation}
  G(i\nu_{n})=\frac{1}{\beta}\sum_{i,j}e^{-i\nu_{n}\tau_{i}^{s}}M_{i,j}
  e^{i\nu_{n}\tau_{j}^{e}}
\end{equation}

Compared with the imaginary time measurement, it seems additional
computational time is needed for the sum over every matrix elements
$M_{i,j}$. K. Haule proposed to implement such measurement in every fast update
procedure which makes sure that only linear amount of time is
needed\cite{Haule-2007}.   

In our calculation the Green's function is measured in the weak-coupling
CT-QMC at each accepted update which greatly reduces the computational
time. The weak-coupling CT-QMC normally yields a higher perturbation order $k$
than the  strong-coupling CT-QMC. It seems that the performance of the
strong-coupling CT-QMC is better\cite{Emanuel-2007}. Concerning the
convergence speed, the weak-coupling CT-QMC is almost same as the
strong-coupling one under the above implementation together with a proper
choice of $\alpha$, since in strong-coupling CT-QMC more Monte Carlo steps are 
needed usually in order to smooth the noise of Green's function at imaginary
time around $\beta/2$ or at large Matsubara frequency points. Furthermore, the
weak-coupling CT-QMC is much easier implemented for large cluster DMFT
calculation, in which case the strong-coupling method needs to handle a big
eigenspace. In this paper we mainly use weak-coupling CT-QMC as impurity 
solver, while all the results can be obtained in the strong-coupling CT-QMC
which was used as an accuracy check.  

Similarly, we adapt K. Haule's implementation to calculate the two-particle
Green's function in frequency space. In the weak coupling CT-QMC, the
non-interacting action has Gaussian form which ensures the applicability of
Wick's theorem for measuring the two particle Green's function 
\begin{eqnarray}\label{2PG}
  \chi_{\sigma\sigma^{\prime}}(\nu_{1},\nu_{2},\nu_{3},\nu_{4})&=&
  T[\overline{G_{\sigma}(\nu_{1},\nu_{2})G_{\sigma^{\prime}}
    (\nu_{3},\nu_{4})}\nonumber\\
  &-&\delta_{\sigma\sigma^{\prime}}\overline{
    G_{\sigma}(\nu_{1},\nu_{4})G_{\sigma}(\nu_{3},\nu_{2})}]
\end{eqnarray} 
The over-line indicates the Monte Carlo average. In each Monte Carlo
measurement, $G(\nu,\nu^{\prime})$ depends on two different argument $\nu$ 
and $\nu^{\prime}$, only in the average level,
$\overline{G(\nu,\nu^{\prime})}=G(\nu)\delta_{\nu,\nu^{\prime}}$ is a function
of single frequency. In each fast-update procedure, the new and old
$G(\nu,\nu^{\prime})$ have a closed relation which ensures that one can
determine the updated Green's function $G^{New}(\nu,\nu^{\prime})$ from the
old one $G^{Old}(\nu,\nu^{\prime})$. For example, adding pair of kinks and 
supposing before updating the perturbation order is $k$, then it is $k+1$
for the new M-matrix. The new inserted pair is at $k+1$ row and $k+1$
column. 
\begin{eqnarray}\label{strong-update}
  &&G^{New}(\nu,\nu^{\prime})-G^{old}(\nu,\nu^{\prime})\nonumber\\
  &=&\frac{M^{New}_{k+1,k+1}}{\beta}G^{0}(\nu)\left\{XL\cdot XR-XR\cdot
    e^{-i\nu\tau_{k+1}^{s}}\right.\nonumber\\
    &&\left.\hspace{0.5cm}-XL\cdot e^{i\nu^{\prime}\tau_{k+1}^{e}}
    +e^{-i\nu\tau_{k+1}^{s}+i\nu^{\prime}\tau_{k+1}^{e}}\right\}
  G^{0}(\nu^{\prime}) 
\end{eqnarray}
Here, $XL=\sum_{i=1}^{k}e^{-i\nu\tau_{i}^{s}}L_{i}$,
$XR=\sum_{j=1}^{k}e^{i\nu^{\prime}\tau_{j}^{e}}R_{j}$ and $L_{i}, R_{j}$ have
the same definition as in Ref\cite{Rubtsov-2005}. In every step, one only needs
to calculates the Green's function when the update is accepted and only a few
calculations are needed. A similar procedure for removing pairs, shiftting
end-point operation can be used. Such method is also applicable in the segment
picture of strong-coupling CT-QMC. In the weak-coupling CT-QMC, such an
implementation greatly improves the calculating speed in low temperature and
strong interaction regime\footnote{In fact, the improvement is more obvious
  for larger M-matrices. The strong coupling CT-QMC and the weak coupling
  CT-QMC require approximately the same amount of CPU time although in the
  weak coupling case the average perturbation order is higher than in the
  strong coupling case}. Once one obtains the two frequency dependent
Green's function in every monte carlo step, the two-particle Green's function
can be determined easily from Eq. (\ref{2PG}). The two-particle vertex is then
given from the following equation:  
\begin{equation}
  \gamma^{\sigma\sigma^{\prime}}_{\omega}(\nu,\nu^{\prime})=
  \frac{\beta^{2}[\chi^{\sigma\sigma^{\prime}}_{\omega}(\nu,\nu^{\prime})
    -\chi^{0}_{\omega}(\nu,\nu^{\prime})]}
  {g_{\sigma}(\nu)g_{\sigma}
    (\nu+\omega)g_{\sigma^{\prime}}(\nu^{\prime}+\omega) 
    g_{\sigma^{\prime}}(\nu^{\prime})}
\end{equation}
where 
\begin{equation}
  \chi^{0}_{\omega}(\nu,\nu^{\prime})=T[\delta_{\omega,0}g_{\sigma}(\nu)
g_{\sigma^{\prime}}(\nu^{\prime})-\delta_{\sigma\sigma^{\prime}}
\delta_{\nu,\nu^{\prime}}g_{\sigma}(\nu)g_{\sigma}(\nu+\omega)]
\end{equation} 
is the bare susceptibility. For the multi-particle Green's function, it still
can be constructed from the two frequency dependent Green's function
$G(\nu,\nu^{\prime})$, but more terms appear from Wicks theorem. Simply, when
set $\nu=\nu^{\prime}$ one can calculate the one-particle Green's funtion
easily. 

\section{Momentum dependece of Vertex}\label{Mom-Vex}

\begin{figure}[t]
  \begin{center}
    \includegraphics[width=230pt]{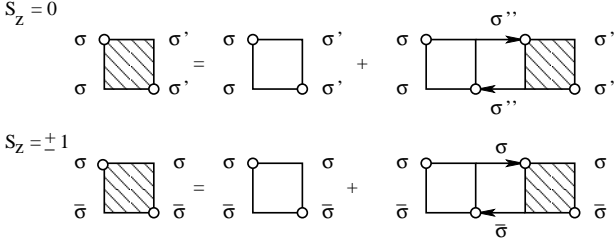}
    \caption{$S_{z}=0$ (ph0) and $S_{z}=\pm1$ (ph1) particle-hole channels of
      the DF vertex, between vertices there are two full DF
      Green's function. The $S_{z}=\pm1$ component is the triplet channel,
      while that for $S_{z}=0$ can be either singlet or triplet.} 
    \label{BSE-channel}
  \end{center}
\end{figure}

As mentioned earlier diagram (a) in Fig. \ref{Self-Energy} only gives the
local contribution. The first non-local correction in the DF method
is from diagram (b). Momentum dependences comes into this theory through the
bubble-like diagram between the two vertices which yields the momentum
dependence of the DF vertex. The natural way to renormalize vertex is
through the Bethe-Salpeter equation. Since the DMFT vertex is only a function
of Matsubara frequency, the integral over internal momentum $k$ and
$k^{\prime}$ ensures that the full vertex only depends on the center of mass
momentum $Q$. The Bethe-Salpeter equation in the particle-hole
channel\cite{Abrikosov:QFT, Nozieres:1964} are shown in
Fig. \ref{BSE-channel}.     

From the construction of the DF method, we know the interaction of the
DF is coming from the two particle vertex of lattice fermion which is
obtained through DMFT calculation. In the Bethe-Salpeter equation, it plays
the role of the building-block. The corresponding Bethe-Salpeter equation for
these two channels are
\begin{subequations}\label{BSE}
  \begin{align}
    & \Gamma^{ph0,\sigma\sigma^{\prime}}_{Q}(\nu,\nu^{\prime}) = 
    \gamma^{\sigma\sigma^{\prime}}_{\omega}(\nu,\nu^{\prime})- \nonumber\\
    &\frac{T}{N}\sum_{k^{\prime\prime}\sigma^{\prime\prime}}
    \gamma^{\sigma\sigma^{\prime\prime}}_{\omega}(\nu,\nu^{\prime\prime})
    G^{d}(k^{\prime\prime})G^{d}(k^{\prime\prime}+Q)
    \Gamma^{ph0,\sigma^{\prime\prime}\sigma^{\prime}}_{Q}
    (\nu^{\prime\prime},\nu^{\prime}) \\
    & \Gamma^{ph1,\sigma\bar{\sigma}}_{Q}(\nu,\nu^{\prime}) = 
    \gamma^{\sigma\bar{\sigma}}_{\omega}(\nu,\nu^{\prime})- \nonumber\\
    &\frac{T}{N}\sum_{k^{\prime\prime}}
    \gamma^{\sigma\bar{\sigma}}_{\omega}(\nu,\nu^{\prime\prime})
    G^{d}(k^{\prime\prime})G^{d}(k^{\prime\prime}+Q)
    \Gamma^{ph1,\sigma\bar{\sigma}}_{Q}(\nu^{\prime\prime},\nu^{\prime})
  \end{align}
\end{subequations} 
Here, the short hand notation of spin configuration is
used. $\gamma^{\sigma\sigma^{\prime}}$ represents
$\gamma^{\sigma\sigma\sigma^{\prime}\sigma^{\prime}}$, while
$\gamma^{\sigma\bar{\sigma}\bar{\sigma}\sigma}$ is denoted by
$\gamma^{\sigma\bar{\sigma}}$ where
$\bar{\sigma}=-\sigma$. $\Gamma^{ph0(ph1)}$ are the full vertices in the
$S_{z}=0$ and $S_{z}=\pm1$ channel, respectively. $G^{d}$ is the full DF
Green's function obtained from section \ref{Details} which is kept unchanged
in the calculation of the Bethe-Salpeter Equation. Different from the work of
S. Brener\cite{Brener-2007}, we solve the above equations directly in momentum
space with the advantage that in this way we can calculate the susceptibility
for any specific center of mass momentum $Q$ and it's convenient to use FFT for
investigating larger lattice.  

In Eq. (\ref{BSE}) one has to sum over the internal spin indices in the
$S_{z}=0$ channel which is not present in $S_{z}=\pm1$ channel. One can
decouple the $S_{z}=0$ channel into the charge and spin channels
$\gamma_{c(s)}=\gamma^{\sigma\sigma}\pm\gamma^{\sigma\bar{\sigma}}$ which can
be solved seperately, and it turns out that the spin channel vertex function 
is exactly same as the that in $S_{z}=\pm1$ channel, see e.g.
P. Nozieres\cite{Nozieres:1964}. Such relation is true for the DMFT vertex,
and was also verified for the momentum dependent vertex in the DF
method\cite{Brener-2007}. In our calculation, we have solved the $S_{z}=0$
channel by decoupling it to the charge and spin channel, while the $ph1$
channel is not used.  

Once the converged momentum dependent DF vertex is obtained, one can
determine the corresponding DF susceptibility in the standard way by
attaching four Green's functions to the DF vertex.  
\begin{subequations}
  \begin{align}
    & \chi^{\sigma\sigma^{\prime}}_{d}(Q) = \chi^{0}_{d}(Q)+\nonumber\\
    & \frac{T^{2}}{N^{2}}
    \sum_{k,k^{\prime}}G^{d}_{\sigma}(k)G^{d}_{\sigma}(k+Q)
    \Gamma^{\sigma\sigma^{\prime}}(Q)
    G^{d}_{\sigma^{\prime}}(k^{\prime})G^{d}_{\sigma^{\prime}}(k^{\prime}+Q) \\
    & \chi^{\sigma\bar{\sigma}}_{d}(Q) = \chi^{0}_{d}(Q)+\nonumber\\
    & \frac{T^{2}}{N^{2}}
    \sum_{k,k^{\prime}}G^{d}_{\sigma}(k)G^{d}_{\bar{\sigma}}(k+Q)
    \Gamma^{\sigma\bar{\sigma}}(Q)G^{d}_{\sigma}(k^{\prime})
    G^{d}_{\bar{\sigma}}(k^{\prime}+Q)
  \end{align}
\end{subequations}
The momentum sum over $\vec{k}$ and $\vec{k}^{\prime}$ can be performed
independently by FFT becasue the DF vertx $\Gamma^{\sigma\sigma^{\prime}}(Q)$
only depends on the center of mass momentum $Q$.

\begin{figure}[t]
  \begin{center}
    \includegraphics[width=230pt]{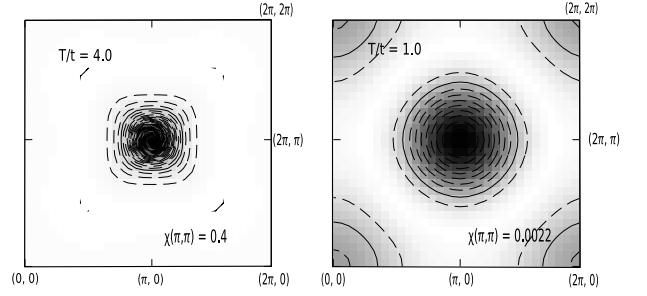}
    \caption{The nontrivial part of the DF spin susceptibilities as a function
      of momentum in 2D Hubbard Model at $U/t=4.0$, $\beta t = 1.0$ (right
      panel) and $\beta t = 4.0$ (left panel). Here 32 $\times$ 32 momentum
      points are used in the first Brillouin zone.}   
    \label{momentum-distribution}
  \end{center}
\end{figure}

Now the z-component DF spin susceptibility $\langle S^{z}\cdot
S^{z}\rangle=\frac{1}{2}(\chi^{\uparrow\uparrow}_{d} 
-\chi^{\uparrow\downarrow}_{d})$ can be determined from the spin channel
component calculated above. In Fig. \ref{momentum-distribution}, 
$\tilde{\chi}^{zz}=\chi^{zz}-\chi_{0}^{zz}$ is shown for $U/t=4$ at
temperatures $\beta t = 4.0$ (left panel) and $\beta t=1.0$ (right
panel). With the lowing down of temperature the DF susceptibility grows up,  
especially at wave vector $(\pi, \pi)$. The momentum $\vec{k}_{x}$ and
$\vec{k}_{y}$ run from $0$ to $2\pi$.  
The susceptibility is strongly peaked at the wave vector $(\pi,\pi)$ at the low
temperature case and the peak value becomes higher and higher. The magnetic
instability of the DF system is indicated by the enhancement of the
DF susceptiblity. The effect of momentum dependence of vertex is clearly
visible in this diagram. The bare vertex which is only a function of frequency
becomes momentum dependent through the Bethe-Salpeter equation. Later on we
will see that such momentum dependent vertex plays a very important role in the
calculation of the lattice fermion susceptibility. 
  
\section{Lattice susceptibility}\label{application} 

The strong antiferromagnetic fluctuation in 2D system is indicated by
the enhancement of the DF susceptibility at the wave vector
$(\pi,\pi)$ shown in Fig \ref{momentum-distribution}. This is the consequence
of the deep relation between the the Green's function of the lattice and the
DF, see  Eq. (\ref{relation}). In order to observe the magnetic
instability of the lattice fermion directly, we have calculated the
lattice susceptibility based on the DF method. By differentiating
the partition function in Eqns. (\ref{original_fermion},
\ref{auxiliary_field}) twice over the kinetic energy, we obtain an exact
relation between the susceptibility of DF and lattice fermions. After some
simplifications\cite{Brener-2007}, it is given by     
\begin{eqnarray}
  && \chi_{f}(Q) = \chi^{0}_{f}(Q) + \nonumber\\
  && \frac{T^{2}}{N^{2}}\sum_{k,k^{\prime}}G^{\prime}(k)G^{\prime}(k+Q)
  \Gamma^{d}_{Q}(\nu,\nu^{\prime})G^{\prime}(k^{\prime})G^{\prime}
  (k^{\prime}+Q)
\end{eqnarray}  
Here $G^{\prime}$ cannt be interpreted as a particle propagator, it is
defined as: 
\begin{equation}
  G^{\prime}(k) = \frac{G^{d}(k)}{g_{\nu}[\Delta_{\nu}-\epsilon(k)]}
\end{equation}
Again, the sum is performed over internal momentum and frequency $k,
k^{\prime}$ which is performed by FFT and rough summing over a few Matsubara
points. Again as in Eq. (\ref{relation}), this equation established a
connection between the lattice susceptibility and the DF
susceptibility. From this point of view, it is easy to understand that the
instability of DFs generates the instability of the lattice
fermions. 

One can also find relations for the higher order Green's function of the
DF and the lattice fermions in the same way. This emphasizes the similar
nature of the DF and lattice fermions except that DF possess only
non-local information, since the DMFT self-consistency ensures that the local
DF Green's function is exactly zero.

The lattice magnetic susceptibility is calculated using the following
definition  
\begin{eqnarray}
  \chi_{m}(q) &=& \frac{1}{N}\sum_{i}e^{iq \cdot
    r_{i}}\int_{0}^{\beta}d\tau e^{-i\omega_{m}\tau}\chi_{f}(i,
    \tau)\nonumber\\ 
  &=& 2(\chi_{f}^{\uparrow\uparrow}-\chi_{f}^{\uparrow\downarrow})
\end{eqnarray}  
where $\chi_{f}(i,
\tau)=\langle
[n_{i,\uparrow}(\tau)-n_{i,\downarrow}(\tau)]\times[n_{0,\uparrow}(0) -
n_{0,\downarrow}(0)]\rangle$. $\chi_{f}$ represents the lattice susceptibility
in order to distinguish with that of the DF. 

\begin{figure}[t]
  \begin{center}
    \includegraphics[width=240pt]{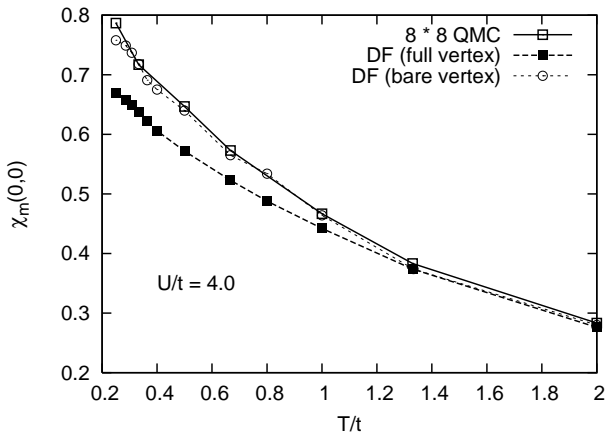}
    \caption{The uniform spin suscetibility of the DF using the bare
      vertex (only frequency dependent) and the full vertex(vertex from the
      Bethe-Salpeter quqation) for half filled 2D Hubbard model at $U/t = 4.0$
      and various temperatures. These results reproduce the similiar solution
      in comparison with the calculation of finite size of
      QMC.}\label{uniform_susceptibility}    
  \end{center}
\end{figure}

We have used two different ways to calculate the lattice susceptibility. First
we have solved the above equation using the bare vertex
$\Gamma(\nu,\nu^{\prime}; \omega)$ which is obtained from the DMFT 
calculation. In contrast, the second calculation was performed using the full
DF vertex. In both of these calculations, the full one particle DF Green's
function was used. The momentum dependent of the DF vertex is
obtained through the calculation of the Bethe-Salpeter equation. The lattice
susceptibility is expected to be improved if we use the momentum
dependence DF vertex. In this way, we can understand the effect of momentum
dependence in the DF vertex.

In Fig. \ref{uniform_susceptibility} we plotted the results for the uniform
susceptibility $\chi_{m=0}(0,0)$ by using both the bare and full DF
vertex. The lattice QMC result\cite{Moreo-1993} is shown for comparison. The
calculation is done for $U/t = 4.0$ and several values of temperature. The
momentum sum is approximated over 32 $\times$ 32 points here. Both of these
calculations reproduce the well known Curie-Weiss law behavior. Surprisingly
enough, the results for the bare vertex fit the QMC results better than that
for the momentum dependent vertex. We believe that this is the finite size
effect of QMC\cite{Moreo-1993}. A. Moreo showed that $\chi$ becomes smaller
when increasing the cluster size $N$. The 4 $\times$ 4 cluster calculation 
result at the same temperature located above of that from 8 $\times$ 8 cluster
calculation. Therefore the results obtained from the full vertex is expected
to be more reliable.  

\begin{figure}[t]
  \begin{center}
    \includegraphics[width=240pt]{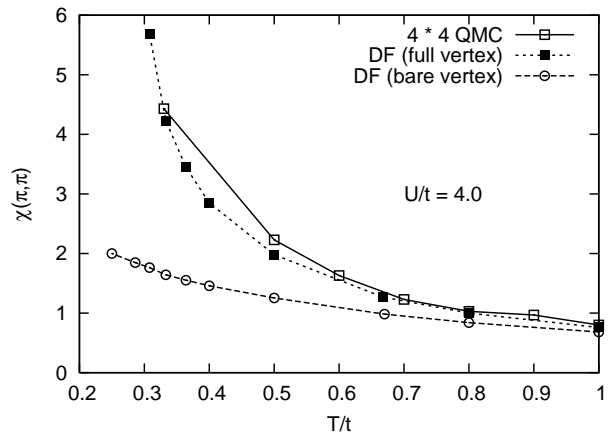}
    \caption{Uniform spin susceptibility at the wave vector $(\pi, \pi)$. The
      QMC results are obtained from Ref.\cite{Bickers-1991(2)}.}
    \label{chi_pi}
  \end{center}
\end{figure}

\begin{figure}[t]
  \begin{center}
    \includegraphics[width=240pt]{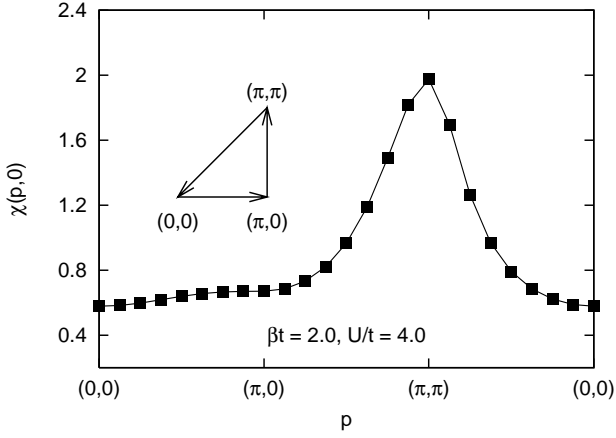}
    \caption{$\chi(q,0)$ vs $q$ at $\beta t= 2.0$, $U/t = 4.0$ for various $q$
    which is along the trajectory shown in the inset.} \label{chi_q} 
  \end{center}
\end{figure}

The importance of the momentum dependence of the DF vertex is more clearly
observed in the calculation of $\chi_{m}(\pi, \pi)$, see
Fig. \ref{chi_pi}. Again, in this diagram QMC  results\cite{Bickers-1991(2)}
are shown for comparison. The same parameters are used as in 
Fig. \ref{uniform_susceptibility}. The result from the DF with bare vertex
does not produce the same results compared with QMC solution. Evenmore
interesting, with decreasing temperature the deviation becomes larger. On the
other hand, the momentum dependent vertex in the DF method gives a
satisfactory answer. This shows the importance of the momentum
dependence in the DF vertex function. Fig.~\ref{chi_q} shows the evolution of
$\chi$ against $q$ for fixed  transfer frequency $\omega_{m}=0$. The path in
momentum space is shown in the inset. From this diagram we can see that
$\chi(q,0)$ reaches its maximum value at wave vector  $(\pi,\pi)$.  

The comparison between the DF and QMC results shows the good
performance of DF method. The DF calculation started from a
single site DMFT calculation and by introducing an auxiliary field, the
non-local information is introduced and nicely reproduces the QMC results. Our
calculation could be done within four hours for each value of the
temperature on average. In this sense, this method is cheap and reliable
compared with the more computationally intensive lattice QMC calculation. 

Similar as the DF method, Dynamical Vertex Approximation
(D$\Gamma$A)\cite{Toschi-2007} is also based on the two particle local
vertex. It deals with the lattice fermion directly, without introducing any
auxiliary field. The perturbative nature of this method ensures its validity
at weak-coupling regime. Unlike in the DF method, D$\Gamma$A takes the
irreducible two particle local vertex as building blocks.  
\begin{subequations}\label{DGA-BSE}
  \begin{align}
    & \gamma_{c(s)}^{-1}(\nu,\nu^{\prime};\omega) = 
    \gamma^{-1}_{c(s),ir}(\nu,\nu^{\prime};\omega) - 
    \chi_{0}(\nu;\omega)\delta_{\nu,\nu^{\prime}} \\
    & \Gamma_{c(s)}^{-1}(\nu,\nu^{\prime};Q) = 
    \gamma^{-1}_{c(s),ir}(\nu,\nu^{\prime};\omega) -
    \chi_{0}(\nu;Q)\delta_{\nu,\nu^{\prime}}
  \end{align}
\end{subequations}
with the spin and charge vertex defined as
$\gamma_{c(s)}=\gamma^{\uparrow\uparrow}\pm\gamma^{\uparrow\downarrow}$. The
bare susceptibility is defined as 
\begin{subequations}
  \begin{align}
    & \chi_{0}(\nu;\omega) = -TG_{loc}(\nu)G_{loc}(\nu+\omega) \\
    & \chi_{0}(\nu,Q) = -\frac{T}{N}\sum_{\vec{k}}
    G^{0}(\vec{k},\nu)G^{0}(\vec{k}+\vec{q},\nu+\omega)
  \end{align}
\end{subequations}
And the self-energy is calculated through the standard Schwinger-Dyson
equation  
\begin{equation}\label{DGA-Selfenergy}
  \Sigma(k) = -U\frac{T^{2}}{N^{2}}\sum_{k^{\prime},Q}
  \Gamma_{f}(k,k^{\prime};Q)G^{0}(k^{\prime})G^{0}(k^{\prime}+Q)G^{0}(k+Q)
\end{equation}
Here, the full vertex $\Gamma_{f}(k,k^{\prime};Q)$ is obtained by summing all
the channel dependent vertices and subtracting the double counted diagrams.
\begin{eqnarray}\label{DGA-FullVertex}
  \Gamma_{f}(k,k^{\prime};Q)&=&\frac{1}{2}\bigg\{
  [3\Gamma_{c}(\nu,\nu^{\prime};Q)-\Gamma_{s}(\nu,\nu^{\prime};Q)]\nonumber\\
  &&-[\Gamma_{c}(\nu,\nu^{\prime};\omega)-
  \Gamma_{s}(\nu,\nu^{\prime};\omega)]\bigg\}
\end{eqnarray}
The one particle propagator is given by the DMFT lattice Green's function
where the self energy is purely local $G^{0}(k)=
1/[i\nu-\epsilon(k)-\Sigma(\nu)]$, the local Green's function is $G_{loc}(\nu)
= 1/[i\nu-\Delta(\nu)-\Sigma(\nu)]$. Then the Dyson equation gives the lattice
Green's function from the self-energy function $G^{-1} = G^{-1}_{0}-\Sigma$. 

\begin{figure}[t]
  \begin{center}
    \includegraphics[width=240pt]{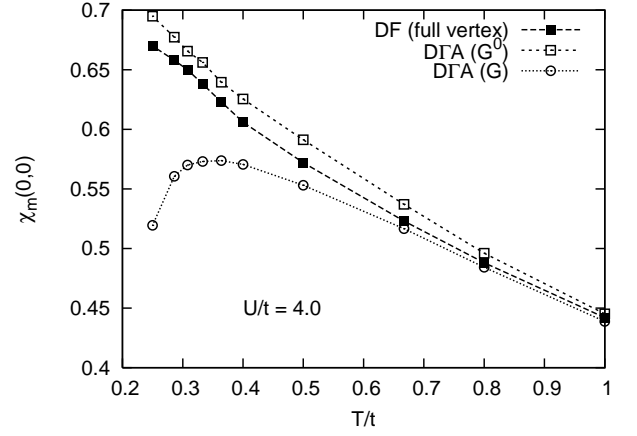}
    \caption{Comparison with the D$\Gamma$A susceptibilities $\chi(0,0)$ which
    obtained from both the DMFT lattice Green's function (D$\Gamma$A
    $(G^{0}$)) and the full Green's function (D$\Gamma$A $(G$)), see context
    for more details.}
  \label{DGA_0}
  \end{center}
\end{figure}

Before presenting the comparison, we take a deeper look at the analysis of 
Eq. (\ref{DGA-BSE}), 
\begin{eqnarray}
  \Gamma_{c(s)}^{-1}(\nu,\nu^{\prime};Q) &=& 
    \gamma^{-1}_{c(s)}(\nu,\nu^{\prime};\omega) - \nonumber \\
    &&[\chi_{0}(\nu;Q)-\chi_{0}(\nu,\omega)]\delta_{\nu,\nu^{\prime}}
\end{eqnarray} 
The second term in the brackets on RHS removes the local term from the
bare susceptibility. The whole term in the brackets then represents only the
non-local bare susceptibility. In order to compare with the DF
method, we take the inverse form of Eq. (\ref{BSE})
\begin{eqnarray}  
  \Gamma_{d,c{s}}^{-1}(\nu,\nu^{\prime};Q) &=&
  \gamma_{c(s)}^{-1}(\nu,\nu^{\prime}, \omega) - \nonumber\\
  && \frac{T}{N}\sum_{\vec{k}}G^{d}(k)G^{d}(k+q) 
\end{eqnarray}
The above two equations are same except for the last term. Since the local
DF Green's function $G^{d}_{loc}$ is zero, the bare DF
susceptibility is purely non-local which coincides with the analysis of
D$\Gamma$A Bethe-Salpeter equation. Therefore, it is not surprising that 
these two methods generate similar results. It is not easy to perform a term
to term comparison between the DF method and D$\Gamma$A although the
bare susceptibilities have no local term in both of these method. The
one particle Green's functions have different meaning in these two methods.  

\begin{figure}[t]
  \begin{center}
    \includegraphics[width=240pt]{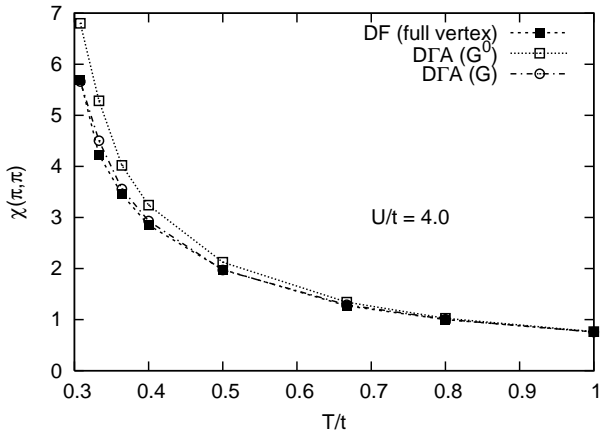}
    \caption{D$\Gamma$A susceptibilities $\chi(\pi,\pi)$ at $U/t=4.0$. The
      susceptibility are determined from both of the DMFT and full lattice
      Green's function together with the vertex obtained from
      Eq. (\ref{DGA-FullVertex})}.  
    \label{DGA_pi}
  \end{center}
\end{figure}

\begin{figure}[t]
  \begin{center}
    \includegraphics[width=220pt]{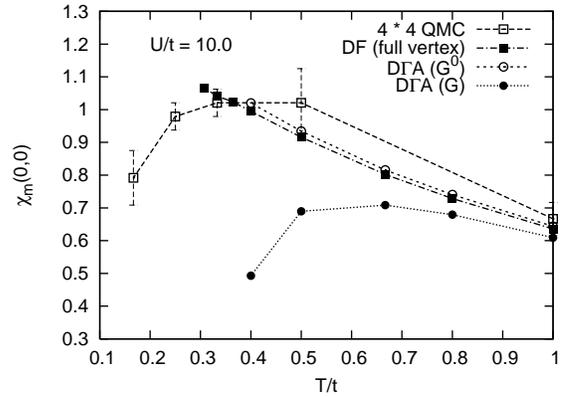}
    \caption{The comparison of the DF resulsts and that of QMC for
      the uniform susceptibility at $U/t=10$. 4$\times$4 QMC
      results\cite{Moreo-1993} also shows the errorbars.}  
    \label{chi_0_U10}
  \end{center}
\end{figure}

The lattice susceptibility within the D$\Gamma$A method is obtained by
attaching four Green's functions on the vertex obtained in
Eq. (\ref{DGA-FullVertex}). There are two possible choices of the lattice
Green's function, one is the DMFT lattice Green's function $G^{0}$, the other
one is the Green's function $G$ constructed by the non-local self-energy from
the Dyson equation. In Fig. \ref{DGA_0} and \ref{DGA_pi}, we presented
the D$\Gamma$A lattice susceptibility calculated from both the DMFT lattice
Green's function labeled as D$\Gamma$A($G^{0}$) and the full Green's function
labeled as D$\Gamma$A($G$). The DF result from the calculation with
the full DF vertex is re-plotted for comparison. In Fig. \ref{DGA_0}, the
D$\Gamma$A susceptibility calculated from the DMFT Green's function
(D$\Gamma$A($G^{0}$)) is basically the same as the DF susceptibility
only with some small deviation. The results for $T/t > 1.0$ which are not shown
here which nicely repeat the DF and QMC results, the deviation
between the D$\Gamma$A and the DF method becomes smaller with the
increasing of temperature. The D$\Gamma$A susceptibility is calculated from
the full Green's function (D$\Gamma$A($G$)) shows a different behavior at low
temperature regime which reached its maximum value at $T/t\approx0.36$. As we
know, the Hubbard Model at half filling with strong coupling maps to the
Heisenberg model, $\chi$ reasches a maximum at $T\approx J$ where 
$J$ is the effective spin coupling constant given as $4t^{2}/U$. The
calculation uses the parameter $U/t=4.0$ which is in the intermediate coupling
regime. Therefore we further calculated the lattice susceptibility at
$U/t=10.0$ which are shown in Fig. \ref{chi_0_U10}. 

When the temperature is greater than 0.4, the DF method and
D$\Gamma$A (D$\Gamma$A($G^{0}$)) generate the similar results to the QMC
calculation. Reducing the temperature further, the QMC susceptibility
greatly drops and shows a peak around 0.4 which coincides with the
behavior of the Heisenberg model. The DF femion and D$\Gamma$A susceptibility
continuously grows up with the decreasing of temperature. Although the
D$\Gamma$A with the full Green's function (D$\Gamma$A($G$)) shows a peak, it
locates at $T/t=0.6667$ which is larger than the peak position of the QMC. And
D$\Gamma$A($G$) generated a large deviation from that of QMC. In 
this diagram, we only show the results of the DF approach for
$T/t>0.3$ and the D$\Gamma$A results for $T/t>0.4$. The Bethe-salpeter
equation of the D$\Gamma$A have a eigenvalue approaching one when further
lowering the temperature, which makes the access of lower temperature region
impossible.    
 
\begin{figure}[b]
  \begin{center}
    \includegraphics[width=220pt]{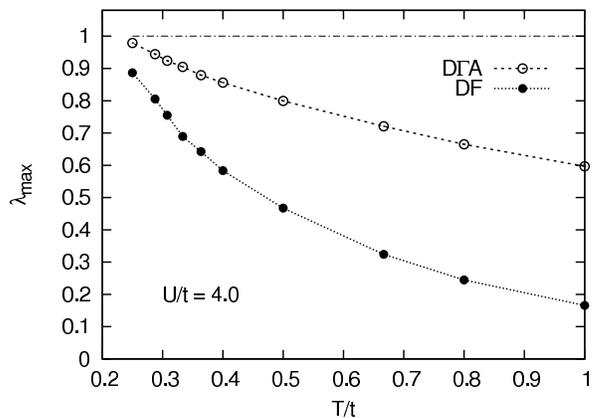}
    \caption{The evolution of maximum eigenvalue in spin channel against
      temperature for DF method and D$\Gamma$A.}
    \label{eigenvalue-T}
  \end{center}
\end{figure}

Fig. \ref{DGA_pi} shows the results of D$\Gamma$A susceptibilities at wave
vector $(\pi, \pi)$. In contrast to the comparison for $\chi(0,0)$ results, the
D$\Gamma$A susceptibility calculated from the full Green's function D$\Gamma$A
($G$) yields better results than that from the calculation with the DMFT
Green's  function D$\Gamma$A ($G^{0}$). D$\Gamma$A ($G$) results are almost on
top of the DF results, the results with DMFT Green's function D$\Gamma$A
($G^{0}$) is large than the DF results. The deviation becomes
larger at lower temperature. Summarizing, the D$\Gamma$A calculation
using the full Green's  function generated the same result as the DF
method for  $\chi(\pi,\pi)$ while failed to produce $\chi(0,0)$ correctly. In
contrast, the calculation with the DMFT Green's function in D$\Gamma$A nicely
produced the results calculated with the DF method for $\chi(0,0)$
while generated larger devivation for $\chi(\pi,\pi)$ at lower temperature
regime compared to that from the DF method. Together with
Fig. \ref{uniform_susceptibility} and \ref{chi_pi}, we can see that the DF
fermion calculation with the full DF vertex generated basically the same
results for both $\chi(0,0)$ and $\chi(\pi,\pi)$ compared to the results of
QMC.  

\begin{figure}[t]
  \begin{center}
    \includegraphics[width=230pt]{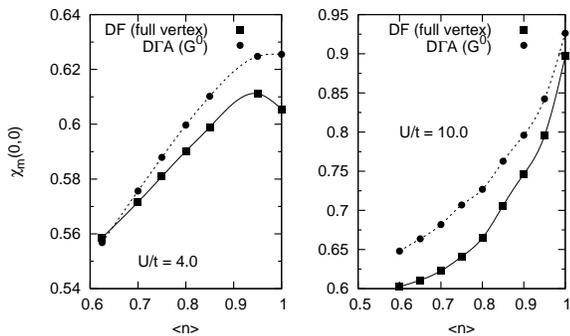}
    \caption{Uniform magentic susceptibility is plotted as a function of
      dopping at $\beta t=2.5$ and $U/t =4.0, 10.0$.}\label{away-half}
  \end{center}
\end{figure}

In both the DF method and the D$\Gamma$A, the operation of inverting
large matrices is required for solving the Bethe-Salpeter
equation. Fig. \ref{eigenvalue-T} shows the leading eigenvalue of 
Eqns. (\ref{BSE}) and  (\ref{DGA-BSE}). As expected, the leading eigenvalue
approaches one with decreasing temperature which directly indicates the
magnetic instability of 2D system. The eigenvalues corresponding to the DF
fermion method always lies  below of that from D$\Gamma$A indicating the
better convergence of the DF method. When the leading eigenvalues are closed to
one, the matrix inversion in Eqns. (\ref{BSE}) and (\ref{DGA-BSE}) are ill
defined, which prevents the investigation at very low temperature. 

Concerning the performance of the DF method, we also calculated the 
uniform susceptibility at away half-filling. In the strong-coupling limit,
the Hubbard model is equivalent to the Heisenberg model with coupling constant
$J=4t^{2}/U$. The consequence of doping is to effectively decrease the
coupling $J$, which yields the increasing behavior of $\chi$ with doping. The
finite size QMC calulation\cite{Moreo-1993, Chen-1993} observed a 
slightly increasing $\chi$ with very small doping at strong interaction
or in the low temperature region. Here, we did a similar calculation at $\beta
t=2.5$ and $U/t=4, 10$. Since the DF method and the D$\Gamma$A do not
suffer from the finite size problem. We would expect to observe results similar
to those of QMC\cite{Moreo-1993,Chen-1993}. In D$\Gamma$A the
suseceptibility is calculated from the DMFT Green's function $G^{0}$  and the
vertex obtained from Eq. (\ref{DGA-FullVertex}). As shown in
Fig. \ref{away-half} at $U/t=4.0$, the susceptibility $\chi$ slightly
increases in the weak dopping region where $\delta$ is around $0.05$, DF
fermion results clearly showed such behavior, D$\Gamma$A also gave a signal of
it. Further doping the system, both the D$\Gamma$A and the DF method
reproduce the decrease with doping as already seen in the QMC. With the
increasing of interaction, we would expect to see the enhancement of this
effect, however our calculation indicates that such increasing-decreasing
behaviro dissappear. Both the D$\Gamma$A and the DF method give the
same decreasing curve which contradict to QMC result\cite{Moreo-1993}. The
results will most likely be further improved by including the higher order
vertex or calculating the cluster DMFT plus DF/D$\Gamma$A\cite{Hafermann-2007}.
    
\section{Conclusion}\label{conclusion}

In this paper, we extended both the DF method and D$\Gamma$A to
calculate the lattice susceptibility. Both of these methods gave equally good
results compared with QMC calculation at $U/t=4.0$. Although they are supposed
to be weak-coupling methods, at $U/t=10.0$ these two methods generated right
results at high temperature region. While both of them failed to reproduce
the Heisenberg physics at low temperature. The investigation of the lattice
susceptibility suffers from hard determined matrix inversion problem at low
temperature regime. The DF methods always generates smaller eigenvalues
compared to D$\Gamma$A indicating the better convergence. The implementation
of DF method in momentum space greatly improves the calculational
speed and makes it easier to deal with larger size lattice. 

\begin{acknowledgments}
We would like to thank the condensed matter group of
A. Lichtenstein at Hamburg University for their hospitality
in particular for the discussions and open exchange
of data with H. Hafermann. Gang Li and Hunpyo Lee would like
to thank Philipp Werner for his help in implementing the
strong-coupling CT-QMC code.
\end{acknowledgments}

\bibliography{chi}

\end{document}